\documentclass[prl,showpacs,superscriptaddress,twocolumn]{revtex4}
\usepackage{dcolumn}
\usepackage{graphicx}
\begin{document}

\raggedbottom

\title{Decelerating molecules with microwave fields}

\author{Simon Merz}
\affiliation{Fritz-Haber-Institut der Max-Planck-Gesellschaft,
Faradayweg 4-6, 14195 Berlin, Germany}

\author{Nicolas Vanhaecke}
\affiliation{Fritz-Haber-Institut der Max-Planck-Gesellschaft,
Faradayweg 4-6, 14195 Berlin, Germany}

\author{Wolfgang J\"{a}ger}
\affiliation{Department of Chemistry, University of Alberta, Edmonton,
Alberta, T6G 2G2, Canada}

\author{Melanie Schnell}
\email[Electronic address: ]{Melanie.Schnell@asg.mpg.de}
\affiliation{Center for Free-Electron Laser Science, 
Notkestrasse 85, 22607 Hamburg, Germany}

\affiliation{Max-Planck-Institut f\"{u}r Kernphysik,
Saupfercheckweg 1, 69117 Heidelberg, Germany}

\author{Gerard Meijer}
\email[Electronic address: ]{meijer@fhi-berlin.mpg.de}
\affiliation{Fritz-Haber-Institut der Max-Planck-Gesellschaft,
Faradayweg 4-6, 14195 Berlin, Germany}

\date{\today}

\begin{abstract}
We here report on the experimental realization of a microwave decelerator
for neutral polar molecules, suitable for decelerating and focusing molecules in 
high-field-seeking states. The multi-stage decelerator consists of a cylindrical 
microwave cavity oscillating on the TE${\it _{11n}}$ mode, with ${\it n}$=12 
electric field maxima along the symmetry axis. By switching the microwave 
field on and off at the appropriate times, a beam of state-selected ammonia 
molecules with an incident mean velocity of 25~m/s is guided while being spatially 
focussed in the transverse direction and bunched in the forward direction. Deceleration 
from 20.0~m/s to 16.9~m/s and acceleration from 20.0~m/s to 22.7~m/s is demonstrated.
\end{abstract}


\maketitle
The manipulation and control of beams of neutral polar molecules with electric
and magnetic fields has enabled spectacular advances during the last decade
and has rejuvenated the field of molecular beams \cite{Meerakker2008,Hogan2011}. 
Beams with a tunable velocity have been used to study (in)elastic scattering as a 
function of collision energy \cite{Scharfenberg2010}, slow molecular beams are being 
used to increase the resolution in spectroscopy experiments \cite{Bethlem2008} and 
samples of trapped molecules have been used to measure lifetimes of long-lived 
metastable states \cite{Meerakker2005}, for instance. These advances have 
greatly benefited from the original development of Stark decelerators \cite{Bethlem1999} 
and Zeeman decelerators \cite{Vanhaecke2007} and their still ongoing refinements. 

The force on a polar molecule can be expressed as the product of an effective dipole 
moment with the gradient of the electric or magnetic field strength. This effective dipole 
moment depends on the quantum state of the molecule and is generally a function of 
the field strength. For the manipulation of molecules, a distinction has to be made 
between molecules in so-called low-field-seeking states and those in high-field-seeking 
states. The electric and magnetic field strength can have a 
minimum in free space and molecules in quantum states with a negative effective 
dipole moment, i.e. molecules in low-field-seeking states, can thus be stably trapped 
with static fields. By creating field minima on the molecular beam axis, these molecules 
can be readily confined around this axis, a feature that is exploited in multipole focusers 
as well as in the Stark and Zeeman decelerators mentioned above. 

There is a strong interest in developing manipulation tools for molecules in high-field-seeking 
states, if only because the absolute ground state of any particle is high field seeking. 
As the Earnshaw theorem does not allow for a maximum of the field strength in free 
space using static fields alone, dynamic focusing (so-called alternating gradient) schemes 
have to be applied to confine molecules in high-field-seeking quantum states. The latter 
schemes are much like those that need to be applied to stably trap charged particles, 
and although guides \cite{Junglen2004,Filsinger2008,Tarbutt2008}, 
decelerators \cite{Bethlem2002} and traps \cite{Veldhoven2005} for neutral molecules 
in high-field-seeking states have all been demonstrated by now, these approaches are 
experimentally more demanding, the confinement is less strong and the trapping is 
intrinsically less stable.

Optical fields do provide a means to create electric field maxima in free space, and are 
used in many laboratories to trap cold atoms as well as cold molecules that have been 
created from cold atoms via photo- or magneto-association \cite{Grimm2000}. Optical fields 
have also been successfully used to decelerate beams of polarizable molecules \cite{Fulton2006}.
The electric field gradients that can be produced in an optical lattice formed by two counter-propagating
pulsed laser beams are very large, leading to extremely rapid deceleration. The volume of 
the optical interaction region is in general rather small, however, and only a limited fraction 
of the molecules in the beam is accepted for the deceleration process. Moreover, it can
be challenging to keep the temporal and spatial intensity profile of the laser pulses sufficiently 
well under control. To circumvent some of these problems, it has been proposed to use 
amplitude modulation of near-resonant infrared radiation inside a high-finesse 
cavity to construct an optical decelerator \cite{Kuma2009}.

A promising alternative for the manipulation of the motion of molecules
is the use of the microwave radiation field inside a resonant cavity, as the volume of the 
interaction region will be much larger and the control over the electric field distribution 
can be more precise. Using high-power microwave radiation, optical traps for polar 
molecules can in principle be made with sub-Kelvin depth \cite{DeMille2004} and a microwave 
decelerator, based on amplitude modulation of the standing wave electric field in an 
open Fabry-Perot microwave cavity, has been proposed \cite{Enomoto2005}. Thus far, 
only a deflector \cite{Hill1975} and a lens \cite{Odashima2010} based on microwaves 
have been experimentally demonstrated for molecules. 

We here demonstrate a microwave decelerator for neutral polar molecules. The 
multi-stage decelerator consists of a cylindrical microwave cavity oscillating on the 
TE${\it _{11n}}$ mode, with ${\it n}$=12 electric field maxima along the symmetry 
axis. A pulsed beam of Stark-decelerated, state-selected ammonia ($^{14}$NH$_3$) 
molecules with an incident mean velocity of around $20-25$~m/s is injected along the axis 
of the cavity. By switching the microwave field on and off at the appropriate times, 
the ammonia molecules are stably guided through the cavity and are focused, 
both transversely and longitudinally, into the detection region, directly behind 
the microwave cavity. Longitudinal spatial focusing ("bunching") of the ammonia 
beam in the detection region, as well as deceleration from 20.0~m/s to 16.9~m/s and 
acceleration from 20.0~m/s to 22.7~m/s are demonstrated.

\begin{figure}[t]
\centering
\includegraphics[width=0.48\textwidth]{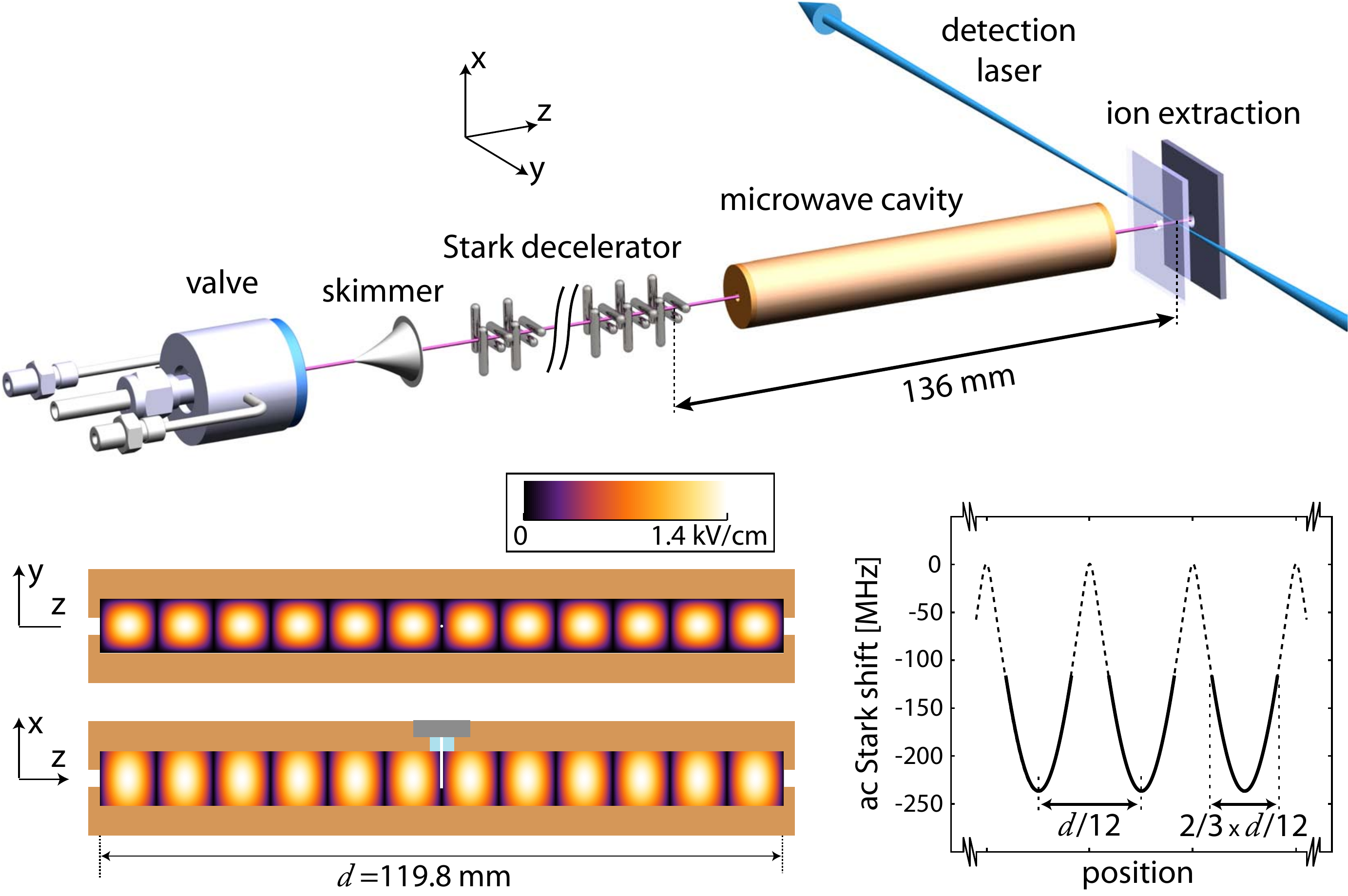}
\caption{(Color online) 
Upper panel: Schematic of the experimental setup. A 12 cm long
cylindrical microwave cavity is mounted behind a Stark decelerator. 
A packet of slow ammonia molecules, exiting the Stark decelerator, 
passes through this cavity and is then detected using resonant photoionization. 
Lower panel: False-color representation of the electric field distribution
inside the microwave cavity when 3.0~W of microwave power at 
$\approx$ 23.7~GHz is coupled into the TE${\it _{1,1,12}}$ mode. 
Cooling the copper cavity to liquid nitrogen temperatures results 
in a Q-factor of about 9150, and a maximum electric field strength 
on the axis of about 1.4~kV/cm. The cavity is represented to scale, but 
for the sake of clarity the antenna is displayed five times thicker than in 
reality. The longitudinal potential wells experienced by the ammonia 
molecules on the axis of the cavity are shown in the lower right corner
for a blue-detuning of the microwave radiation of 58~MHz from the 
molecular resonance. The part of the potentials experienced by the 
synchronous molecule in guiding mode is represented in bold.
}
\label{fig:ExperimentalSetUp}
\end{figure}

The experimental setup is schematically shown in Figure \ref{fig:ExperimentalSetUp}.
A detailed description of the molecular beam machine and, in particular, of the
Stark deceleration of a beam of ammonia molecules is given elsewhere \cite{Bethlem2002b}.
In the experiment, a single decelerated packet of NH$_3$ molecules leaves the
Stark decelerator in the upper inversion doublet component of the 
$\mid$J,K$\rangle$=$\mid$1,1$\rangle$ level, in the vibrational and electronic ground state.
The mean longitudinal velocity of the molecules is set around $20-25$~m/s and the
full-width-at-half-maximum (FWHM) velocity spread is about 10~m/s. The 
decelerated packet leaving the decelerator contains about 10$^5$ molecules and has a spatial
extent of less than 2~mm along all directions. We here define the time at which
the Stark decelerator is switched off as $t$=0. At this time, the center of the packet of 
molecules is about 6~mm in front of the microwave cavity. The inner length of the microwave 
cavity is $d$=119.8~mm (at room temperature), and there are identical 2~mm thick end-caps 
(with a 3~mm diameter opening for the molecules) on either end. About 6~mm behind the 
microwave cavity the ammonia molecules are resonantly ionized between the extraction 
plates of a compact linear time-of-flight setup and the parent ion signal is recorded. 
The total distance that the molecules travel from $t$=0 up to the time of detection is thus 
about 136~mm. 

The cylindrical microwave cavity is made out of oxygen-free copper and has a precisely 
machined inner diameter of 9.57~mm (at room temperature). Microwave radiation around 
23.7~GHz is coupled in via a 6.5~mm long, 0.1~mm diameter dipole antenna, located at the 
center of the cavity, to excite the TE${\it _{1,1,12}}$ 
mode. When the cavity is cooled down to 77~K, this cavity mode is about 58~MHz 
blue-shifted from the molecular resonance frequency $\nu_0$ $\approx$ 23695 MHz, 
i.e. from the inversion transition of the $\mid$J,K$\rangle$=$\mid$1,1$\rangle$ level 
of the vibronic ground state of $^{14}$NH$_3$. From the observed width of the cavity 
resonance, a quality factor of the resonator of Q=9150 is deduced. Due to the blue-detuning, 
molecules in the upper component of the inversion doublet, that are low-field seeking in static 
electric fields, become high-field seeking in the microwave field \cite{Odashima2010}. 
The radial dependence of the electric field in the microwave cavity, which is always greatest 
on the beam axis, thus results in a focusing of the ammonia molecules along this axis. 
This behavior was exploited in the demonstration of a microwave lens for ammonia molecules, 
for which low order TE${\it _{11n}}$ modes with $n$ =2 or 4 were used \cite{Odashima2010}. 

The mode with $n$=12 electric field maxima on the symmetry axis that is used here, 
creates electric field gradients in the longitudinal direction that are similar in magnitude 
to those in the transverse direction, as can be seen from the mode pattern shown in the 
lower left corner of Figure \ref{fig:ExperimentalSetUp}.
The longitudinal gradients can be used to influence and control the forward velocity of the molecules.
When 3.0~W of microwave power is coupled into the cavity, the maximum electric field 
strength on the axis is around 1.4~kV/cm, resulting in an ac Stark shift of approximately 
250~MHz. As this is considerably larger than the detuning of 58~MHz, the Stark shift that 
the molecules experience near the field maxima scales almost linearly with the electric field 
strength \cite{Odashima2010}. This in turn implies that the longitudinal potential that 
the ammonia molecules experience is nearly perfectly harmonic around the position of each 
of the twelve electric field maxima, as shown explicitly in the lower right corner of 
Figure \ref{fig:ExperimentalSetUp}.

\begin{figure}[tb] \centering
\includegraphics[width=0.36\textwidth]{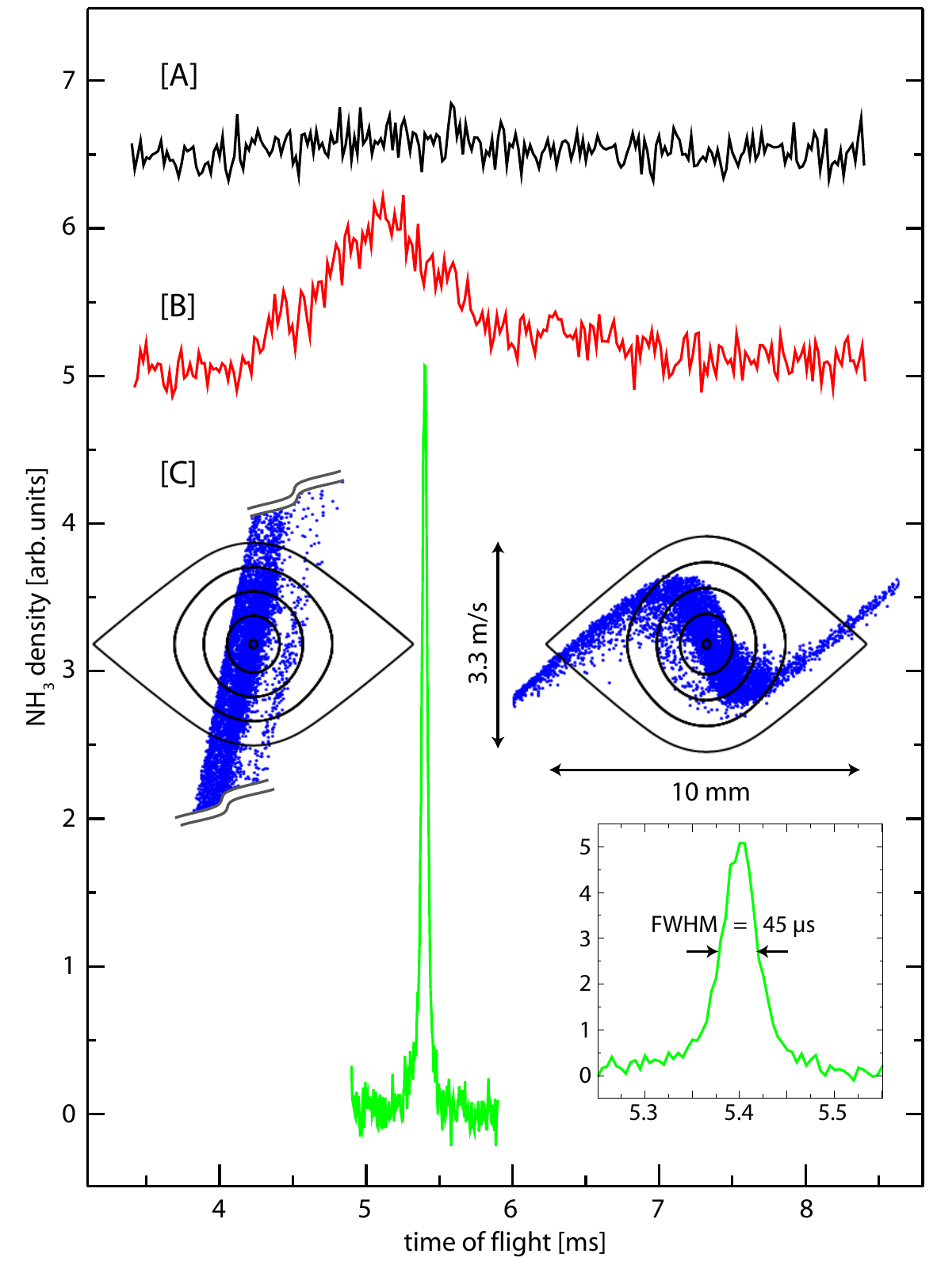}
\caption{(Color online)
Density of ammonia molecules behind the microwave cavity as a function 
of time (in milliseconds) without microwaves ([A]), with 3.0~W of microwaves continuously
coupled in ([B]) or with pulsed application of the microwaves for two thirds of the time ([C]). 
The curves are offset for clarity, but all have the same vertical scale. For the measurements 
in panel [C], the microwaves are applied synchronously with the 
motion of an ammonia molecule that moves on the molecular beam axis with a forward 
velocity of 25~m/s. The longitudinal phase space distribution of the molecules at the
first time that the microwaves are switched on and at the last time that they are
switched off is shown on the left and right side of the peak, respectively, overlaid on the
longitudinal phase space acceptance diagram.
}
\label{fig:MicrowaveGuiding}
\end{figure}

Figure \ref{fig:MicrowaveGuiding} shows the density of ammonia molecules 
behind the microwave cavity as a function of time. Without microwave radiation 
in the cavity (panel [A]), there is hardly any detectable signal; the slow, divergent beam 
of ammonia molecules is strongly diluted during free flight to the laser interaction 
region. When
3.0~W of microwaves are continuously coupled in (panel [B]), the cavity functions as a 
microwave lens and the ammonia molecules are transversally focused into the
detection region. The observed broad arrival time distribution, centered around 
5.2~ms, reflects the broad initial velocity distribution of the molecules entering the 
microwave cavity. In panel [C], measurements are shown that are recorded using 
a longitudinal guiding mode of operation. The microwaves are switched on and off 
every time a (fictitious) molecule that enters the cavity on the beam axis with a 
forward velocity of $v_z$=25~m/s reaches the position at which the electric field 
strength is 50\% of its maximum value. This means that the microwave fields are 
on for a distance spanning two thirds of each electric field strength period
($2/3 \times d/12 \approx 6.7$~mm), 
symmetrically spread around the electric field maxima, as schematically shown in 
the lower right corner of Figure \ref{fig:ExperimentalSetUp}. Consequently, during its 
flight on the axis of the cavity, such a molecule, referred to as the synchronous 
molecule, only experiences the deepest part of the potential wells. The ammonia 
molecules are thus trapped in an effective traveling longitudinal potential well, 
much like the situation during guiding at a constant velocity in a Stark decelerator
\cite{Bethlem2000}. 

The observed arrival time distribution in panel [C] has an intense peak
around $t$=5.40~ms, the expected arrival time for molecules being transported 
through the microwave cavity with a mean velocity of 25~m/s. This narrow peak is shown
on an expanded time-scale in the lower right corner of Figure \ref{fig:MicrowaveGuiding},
from which it is seen to have a temporal width of 45~$\mu$s. This width mainly results 
from the length of the packet of molecules, although there might be a slight broadening 
due to the finite size of the (focused) ionization laser beam. Neglecting the latter 
effect sets the upper limit on the longitudinal size of the packet of ammonia molecules 
in the detection region to only 1.1~mm.
The packet is this compact because we selected the parameters such as to perform 
longitudinal spatial focusing in the detection region, i.e. a one-to-one 
image is made of the packet at the exit of the Stark decelerator in the detection 
region. The molecules contributing to this narrow peak would have given rise to a ten
times broader time of flight profile when the microwaves would have been kept off,
or constantly on.

The principle of bunching is visualized in the insets to panel [C], where 
the calculated longitudinal phase space distributions of the molecules near the entrance 
(left side) and close to the exit (right side) of the microwave cavity are 
shown, overlaid on the longitudinal phase space acceptance of the 
microwave decelerator. Under the present operating conditions, the packet 
undergoes almost exactly half a rotation in phase space, such that a spatially
narrow packet, but with a broad longitudinal velocity distribution, is created in 
the detection region; a narrow longitudinal velocity distribution, at the cost of 
an extended spatial distribution, can be created if the parameters are selected 
to perform only a quarter of a full rotation in phase space in the microwave cavity.
These bunching and longitudinal cooling schemes in a microwave cavity for 
molecules in high-field-seeking states are similar to schemes demonstrated 
before with switched electric fields for molecules in low-field-seeking states \cite{Crompvoets2002}.

\begin{figure}[htpb]
\centering
\includegraphics[width=0.355\textwidth]{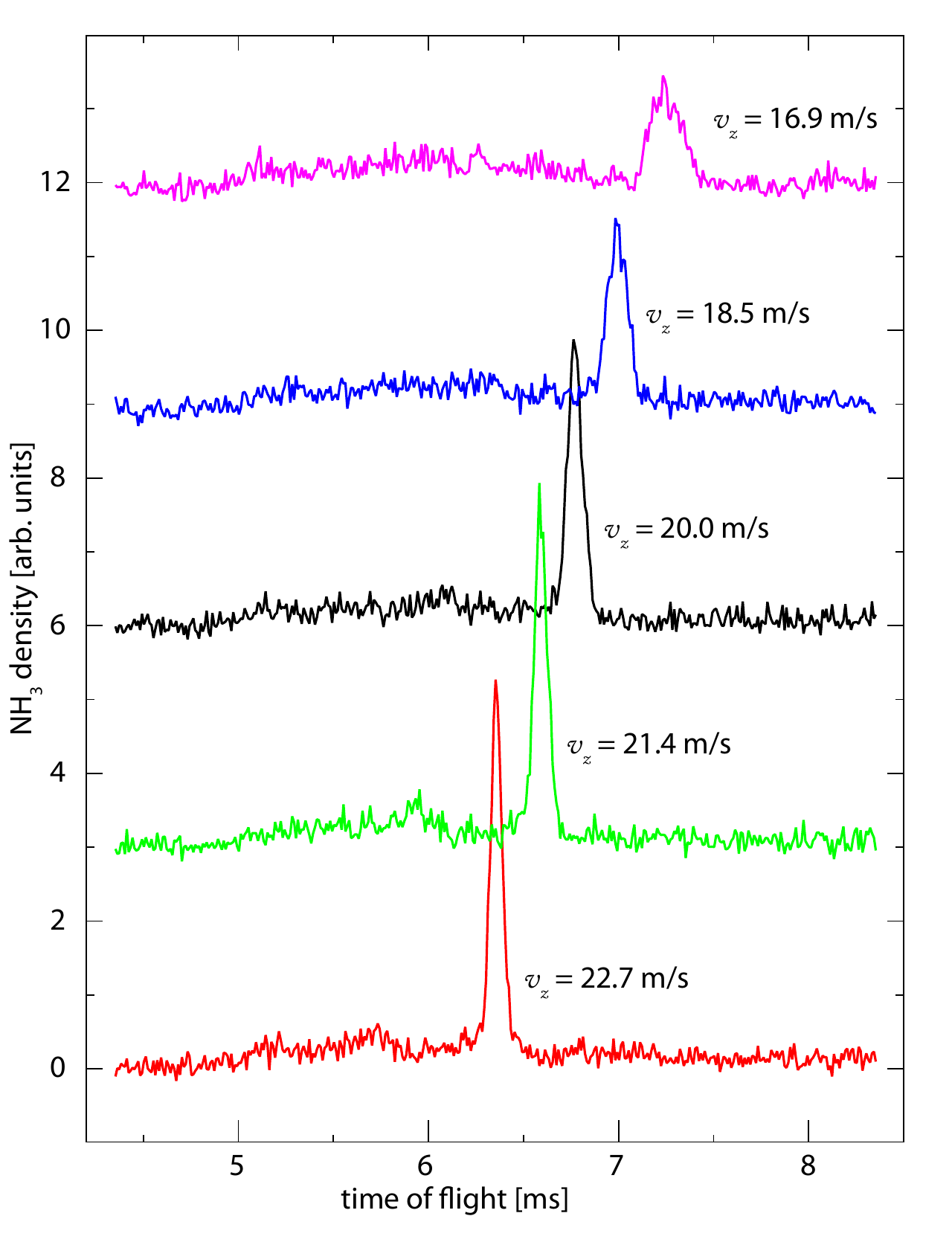}
\caption{(Color online)
Density of ammonia molecules behind the microwave cavity as a function 
of time for two different acceleration, a guiding and two deceleration sequences 
(curves vertically offset for clarity). Ammonia molecules that enter the cavity 
with $v_z$=20.0~m/s exit with the indicated velocity.
}
\label{fig:MicrowaveDeceleration}
\end{figure}

The results of microwave acceleration and deceleration experiments are shown in 
Figure \ref{fig:MicrowaveDeceleration}. For these measurements, the microwaves 
are again switched on and off such that the synchronous molecule experiences two 
thirds of the potential wells, but in intervals that are no longer distributed symmetrically 
around the electric field maxima. In addition, the corresponding time intervals are no 
longer equidistant, due to the non-zero acceleration. The microwave fields are switched 
such that a synchronous ammonia molecule that enters the microwave cavity with an 
initial velocity of $v_z$=20.0~m/s gains or loses a fixed amount of kinetic energy 
$h \times$100~MHz ($h \times$201~MHz) per microwave deceleration stage, where 
$h$ is Planck's constant. These molecules then exit the cavity with velocities 
of 21.4~m/s (22.7~m/s) or 18.5~m/s (16.9~m/s), respectively, as indicated next to the 
observed arrival time distributions. The time of flight distribution obtained when guiding 
molecules at a constant velocity of 20.0~m/s is shown as the middle trace. Although the longitudinal 
acceptance of the microwave decelerator gets smaller with increasing acceleration 
and deceleration rates, the peak intensity in the arrival time distributions is seen to get 
slightly higher for the accelerated beams. This is explained by the improved transverse 
and longitudinal spatial focusing in the detection region in this case.  

In conclusion, a prototype microwave decelerator, enabling excellent control over the 
six-dimensional phase space distribution of an ensemble of molecules in 
high-field-seeking states, has been experimentally demonstrated. This microwave 
decelerator can be used, for instance, to prepare samples of ground-state molecules 
for subsequent spectroscopy and scattering experiments or to load them into 
a (microwave) trap. 

\begin{acknowledgments}
We acknowledge help in the design of the cooled microwave 
cavity from H. Haak and useful discussions with B.~G. Sartakov.  This 
work has been supported by the ERC-2009-AdG program under grant agreement 
247142-MolChip. W.J. gratefully acknowledges the support of the Alexander 
von Humboldt Foundation, and M.S. the support of the Fonds der Chemischen Industrie.
\end{acknowledgments}

\end{document}